\documentclass[aps,prl,preprintnumbers,amsmath,amssymb,superscriptaddress]{revtex4}%

\usepackage{graphicx}%
\usepackage{dcolumn}
\usepackage{amsmath}
\usepackage{amssymb}
\usepackage{color}
\usepackage{multirow}

\newcommand{\cis}{Ca$_{3}$\-Ir$_{4}$\-Sn$_{13}$ }
\newcommand {\sis}{Sr$_{3}$\-Ir$_{4}$\-Sn$_{13}$ }
\newcommand {\csis} {(Ca$_{1-x}$Sr$_{x}$)$_{3}$\-Ir$_{4}$\-Sn$_{13}$ }
\newcommand{\mSR}{$\mu$SR }
\newcommand{\Tc}{$T_{\rm c}$ }

\begin{document}


\title {Strong enhancement of s-wave superconductivity near a quantum critical point of Ca$_3$Ir$_4$Sn$_{13}$}

\author{P.~K.~Biswas}
\affiliation{Laboratory for Muon Spin Spectroscopy, Paul Scherrer Institute, CH-5232 Villigen PSI, Switzerland}
\author{Z. Guguchia}
\affiliation{Laboratory for Muon Spin Spectroscopy, Paul Scherrer Institute, CH-5232 Villigen PSI, Switzerland}
\author{R.~Khasanov}
\affiliation{Laboratory for Muon Spin Spectroscopy, Paul Scherrer Institute, CH-5232 Villigen PSI, Switzerland}
\author{M.~Chinotti}
\altaffiliation[Current address: ]{Laboratory for Solid State Physics, ETH Zurich, CH-8093 Zurich, Switzerland}
\affiliation{Laboratory for Muon Spin Spectroscopy, Paul Scherrer Institute, CH-5232 Villigen PSI, Switzerland}
\author{L. Li}
\altaffiliation[Current address: ]{Key Laboratory of Materials Physics, Institute of Solid State Physics, Chinese Academy of Sciences, Hefei 230031, China}
\affiliation{Condensed Matter Physics and Materials Science Department, Brookhaven National Laboratory, Upton, New York 11973, USA}
\author{Kefeng~Wang}
\altaffiliation[Current address: ]{Department of Physics, University of Maryland, College Park, Maryland 20742-4111, USA}
\affiliation{Condensed Matter Physics and Materials Science Department, Brookhaven National Laboratory, Upton, New York 11973, USA}
\author{C.~Petrovic}
\affiliation{Condensed Matter Physics and Materials Science Department, Brookhaven National Laboratory, Upton, New York 11973, USA}
\author{E.~Morenzoni}
\email[Corresponding author: ]{elvezio.morenzoni@psi.ch}
\affiliation{Laboratory for Muon Spin Spectroscopy, Paul Scherrer Institute, CH-5232 Villigen PSI, Switzerland}
\date{\today}


\begin{abstract}


We report microscopic studies by muon spin rotation/relaxation as a function of pressure of the ~\cis and \sis cubic compounds, which are members of the \csis system displaying superconductivity and a structural phase transition associated with the formation of a
charge density wave (CDW). We find a strong enhancement of the superfluid density and a dramatic increase of the pairing strength above a  pressure of $\approx 1.6 $ GPa giving direct evidence of the presence of a quantum critical point separating a superconducting phase coexisting with CDW from a pure superconducting phase.
The superconducting order parameter in both phases has the same $s$-wave symmetry.
In spite of the conventional phonon-mediated BCS character of the weakly correlated \csis system, the dependence of the effective superfluid density on the critical temperature puts this compound in the ``Uemura'' plot close to unconventional superconductors. This system exemplifies that conventional BCS superconductors in the presence of competing orders or multi-band structure can also display characteristics of unconventional superconductors.


\end{abstract}

\maketitle

\section{Introduction}

The interplay between different electronic ground states is one of the fundamental topics in condensed matter physics and is well apparent
in phase diagrams as a function of doping, pressure or magnetic fields, resulting in various forms of coexistence, cooperation or competition of the order parameters \cite{Keimer,Paglione,Scalapino,Fradkin}.
Particularly interesting are the regions at phase boundaries or at quantum critical points (QCPs) where different quantum states meet \cite{Coleman}.
Very often magnetism and superconductivity are involved and, in spite of diverse structural and physical properties, many compounds show characteristic phase diagrams where superconductivity is found in the vicinity of electronic instabilities of magnetic (mainly antiferromagnetic) origin. In this case spin fluctuations are predominantly considered at the heart of the mechanisms leading to pairing
and superconductivity is unconventional.
Less common is the case where the electronic instability is linked to the formation of a charge density wave (CDW), which is based on the same electron-phonon interaction found in conventional superconductors.

Ternary intermetallic stannide compounds such as $R_3T_4$Sn$_{13}$, where $R$ =La, Ca, Sr and $T$=Ir, Rh  \cite{Remeika,Espinosa} are of particular interest because they exhibit many physical properties such as superconductivity, magnetic or charge order, and structural instabilities.
The quasiskutteridite cubic superconductor (Ca,Sr)$_3$Ir$_4$Sn$_{13}$ and the related (Ca,Sr)$_3$Rh$_4$Sn$_{13}$ have recently attracted attention because of the presence of a  pressure induced structural phase transition at a temperature $T^*$, the possible coexistence of superconducting and charge density wave states, and a putative quantum critical point ~\cite{Klintberg,Zhou,Mazzone,Kuo14,Wang15,Wang,Fang14,Goh15,Kuo15}.
The role and interplay of these degrees of freedom remain a central issue also in many unconventional superconductors, as demonstrated by the recent observation of
CDW in HgBa$_2$CuO$_{4+\delta}$  \cite{Tabis} and of competition between superconductivity and charge order in YBa$_2$Cu$_3$O$_{6.67}$ \cite{Chang}.
In (Ca$_x$Sr$_{1-x}$)$_3$Ir$_4$Sn$_{13}$, the phase transition was found following the observation of an anomaly in temperature dependent resistivity and susceptibility measurements at $T^\ast \simeq{147}$~K and $T^*\simeq{33}$~K in Sr$_3$Ir$_4$Sn$_{13}$ and Ca$_3$Ir$_4$Sn$_{13}$, respectively. Initially, the anomaly was attributed to ferromagnetic spin fluctuations, coexisting and possibly enhancing the superconductivity appearing at lower temperature $T_{\rm c}\simeq{5}$~K and $T_{\rm c}\simeq{7}$~K in Sr$_3$Ir$_4$Sn$_{13}$ and Ca$_3$Ir$_4$Sn$_{13}$, respectively ~\cite{Yang}. Later, single crystal x-ray diffraction and neutron scattering studies showed that the  anomaly in Sr$_3$Ir$_4$Sn$_{13}$ and \cis is produced by a second-order structural transition at $T^*$ \cite{Klintberg, Mazzone}.

Various physical quantities from transport and magnetization measurements suggest that the transition is associated with a charge density wave transition involving the conduction electrons system ~\cite{Klintberg}.
Measurements of Hall and Seebeck coefficients and $^{119}$Sn NMR indicate decrease of the carrier density and significant Fermi surface reconstruction with change of sign of the Hall coefficient at $T^*$ in Ca$_3$Ir$_4$Sn$_{13}$ and Sr$_3$Ir$_4$Sn$_{13}$ ~\cite{Wang,Kuo14,Wang15}.
Optical spectroscopy reveals the formation of a partial energy gap at the Fermi surface associated with the structural phase transition \cite{Fang14}.

$\mu$SR measurements of the two end compounds \cis and the isoelectronic sister compound \sis at ambient pressure did not find any sign of magnetic ordering or weak magnetism of static or dynamic origin \cite{Gerber, Biswas,Biswas14b} and found BCS-like superconductivity with $s$-wave symmetry and a $\Delta(0) / (k_{\mathrm{B}} T_c)$ ratio typical of strong coupling similar to other $R_3T_4$Sn$_{13}$ compounds \cite{Kase11,Hayamizu}.
Since the atomic size of Ca is smaller than that of Sr, the substitution of Ca on the Sr site corresponds to applying a positive pressure of about 5.2 GPa, which
reduces $T^*$. This behavior continues in \csis for external pressure. At the same time an increase of $T_{\rm c}$ with increasing pressure has been observed so that, under hydrostatic pressure, the $T_{\rm c}$ of Ca$_3$Ir$_4$Sn$_{13}$ reaches 8.9~K at $\sim$ 4.0 GPa and then falls for higher pressures \cite{Klintberg}. A linear extrapolation of the $T^*(p)$ dependence to $T^*$=0 predicts a structural/CDW quantum critical point at $\approx 1.8$ GPa. Note, however, that this value has been extracted from measurements of $T^*$ in the normal state, so that data available up to now do not give direct evidence of a QCP in the superconducting state.


The increase of $T_{\rm c}$ with pressure and the concomitant suppression of $T^*$ point to an intimate interplay between CDW and superconductivity, possibly culminating in a quantum critical point where the CDW is completely suppressed. There is a long-standing question concerning the competition, coexistence or cooperation between the two electronic states as well as about the role of charge density fluctuations.
A CDW opens a gap on part of the Fermi surface and can therefore profoundly modify the microscopic properties and gap structure of the superconductor.
The \csis compound allows to investigate the questions of competing electronic/structural instabilities, as well as whether these states are separated by a QCP and how the pairing state evolves around such a point.

Here, we use the $\mu$SR technique to characterize at a microscopic level the superconducting and magnetic properties in \cis and \sis as a function of pressure.
In particular, from measurements of the inhomogeneous field distribution in the vortex state $p(B)$ we determine the effective magnetic penetration depth $\lambda(T)$ and its temperature and pressure dependence.
$\mu$SR is a powerful tool to determine the absolute value of $\lambda(T)$ and its dependence on various thermodynamic parameters.
The magnetic penetration depth is a fundamental property of the superconducting state, that can be measured to probe the electronic
structure of the material and to look for signatures of a QCP.
The temperature dependence of $\lambda^{-2}$ is a measure of the superfluid density $\lambda^{-2} \propto  \rho_s \equiv \frac{n_s}{m^*}$ (where $n_s$ is the supercarrier density and ${m^*}$ effective mass), whereas
the low-temperature behavior of $\lambda(T)$ reflects the superconducting gap structure.
Despite the importance of the evolution of this quantity in understanding the nature of superconductivity, especially around a zero temperature quantum transition, there are not many measurements on the pressure dependence
on $\lambda$ and only a small number of $\mu$SR studies on the effect of pressure have been reported so far \cite{Khasanov04,Khasanov05,Maisuradze11}.

Here we observe a pronounced increase of the superfluid density that sets in at a pressure $p_c \approx 1.6$ GPa with a sudden enhancement of the superconducting gap
value, indicating that a QCP exist in the superconducting state leading to a significant enhancement of the superconducting pairing strength when the CDW is suppressed.
We find that \cis remains a conventional electron-phonon coupled $s$-wave superconductor across the quantum phase transition. The strenghtening of the superconducting coupling at $p_c$ appears to be related to the phonon softening at the structural quantum critical point. Although a conventional superconductor, the dependence of \Tc on superfluid density resembles that of unconventional superconductors when plotted in the so called ``Uemura'' plot. This system exemplifies that conventional BCS superconductors can also display unconventional features in the presence of competing orders or multiband structure.

\section{Results}\label{sec:Results}

Single crystal samples of Ca$_3$Ir$_4$Sn$_{13}$ and \sis were grown as described in Ref. \cite{Wang}.
Transverse-field (TF) $\mu$SR experiments were performed on the GPD instrument at the $\mu$μE1 beam line of the Paul Scherrer Institute (Villigen, Switzerland).
High-energy muons (p$_\mu \sim$ 100 MeV/c) were implanted in the sample. Forward and backward positron detectors with respect to the initial muon polarization were used for the measurements of the $\mu$SR asymmetry time spectrum $A(t)$.
The spectra yield information about the superconducting properties. For instance, the average muon spin precession frequency is directly proportional to the average local magnetic field at the muon site and can detect a diamagnetic shift associated with the supercurrents. In the vortex state with a typical field distribution, the precession is damped and the damping is related to the field distribution, which depends on the characteristic length scales of the superconductor \cite{SM}.

Muon spin rotation (\mSR\!) spectra in the vortex state were taken as a function of increasing temperature by initially
field-cooling the sample down to 0.3 K in a 50 mT in the case of \cis and 30 mT field for \sis and for different pressures.
Typical statistics for a $\mu$SR spectrum were about 5 $\times$ 10$^6$ positron events in the forward and backward detectors.  Several (single crystal) pieces of the compound were loaded into the cylindrical pressure cell. The sample dimensions were chosen to maximize the filling factor of the pressure cell (diameter 6 mm, height 15 mm). A CuBe piston-cylinder pressure cell was used with Daphne oil as
a pressure-transmitting medium for pressures up to 1.1 GPa; from 1.1 GPa and above MP35N was used. The maximum pressure achieved at low temperature is about 2.2 GPa. The pressure was calibrated
by measuring via AC susceptibility the superconducting transition of a very small indium plate inserted in the cell.
The pressure point at 1.1 GPa was measured with the two different cells to better assess the background contribution arising from muons stopping in the cell walls and check that the use of the two cell materials has no effect on the results. The fraction of the muons stopping in the sample was approximately 50\% with the CuBe cell and 40\% with the MP35N cell. Resistivity and magnetoresistance measurements to determine $\rho(T)$ and $B_{c2}(T)$ were performed using a PPMS \textit{Quantum Design} instrument. Additional characterization of the samples grown under the same conditions, such as specific heat, Seebeck coefficient and Hall resistivity, can be found in Ref. \cite{Wang}.

\begin{center}
\begin{figure}[htb]
\includegraphics[width=0.9\linewidth]{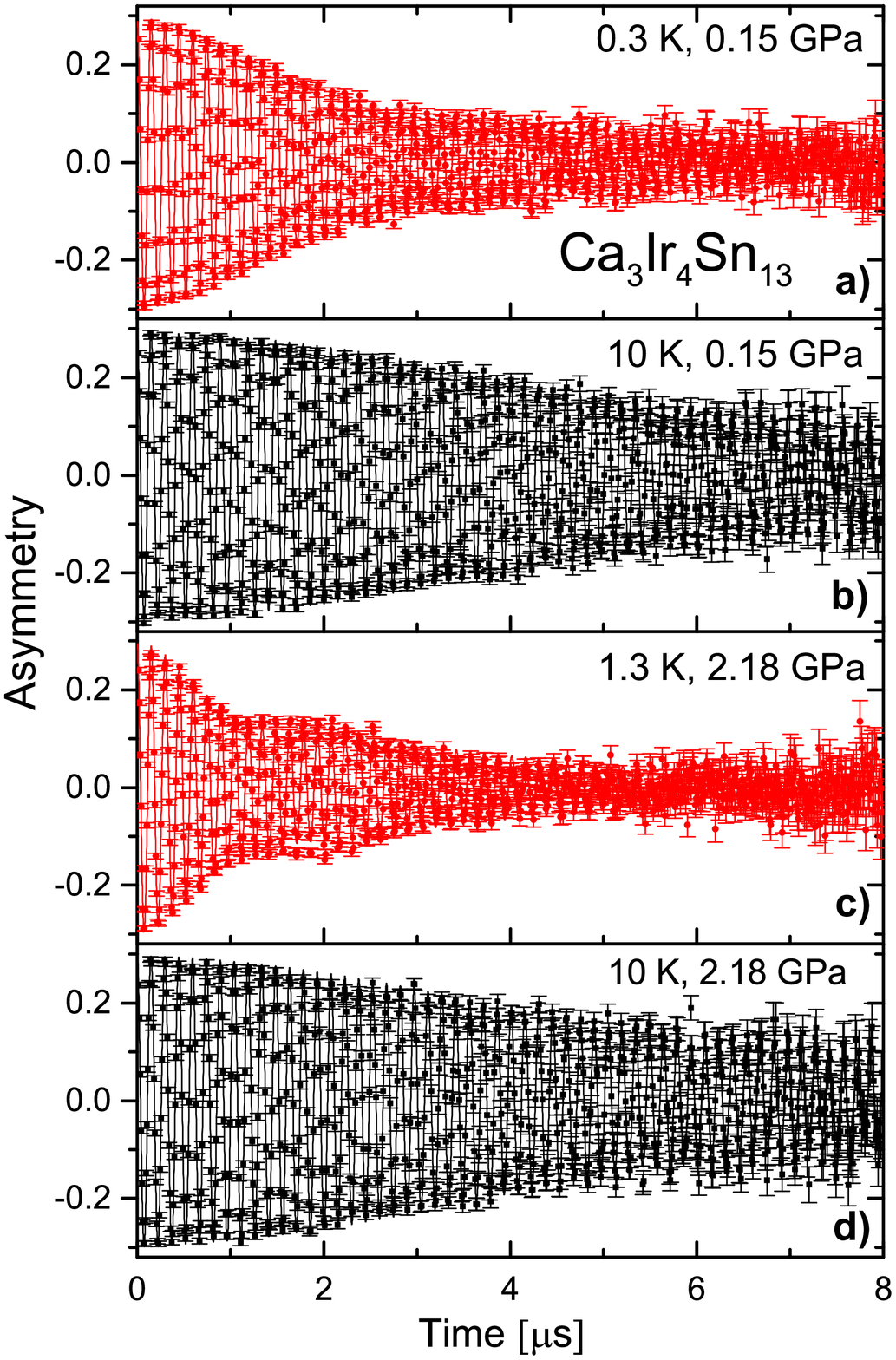}
\vspace{0cm}
\caption{Asymmetry spectra for Ca$_{3}$Ir$_{4}$Sn$_{13}$ measured below (figures a) and c)) and above (figures b) and d)) $T_{c}$ and at pressures of 0.15 GPa (figures a) and b)) and 2.18 GPa (figures c) and d)) after cooling the sample in a transverse field (TF) of 50 mT. The increased damping below \Tc at shorter times reflects the formation of a vortex lattice and the dephasing of the muon ensamble in the corresponding field distribution. At higher pressure the damping is more pronounced reflecting a shorter magnetic penetration depth. The residual damping observed in the normal state is due to the contribution of the nuclear moments in the sample. The solid line are the fits, as described in the Supplemental Material \cite{SM}, which take also into account the background contribution from the muons stopping in the pressure cell walls.}
\label{fig:msr spectra}
\end{figure}
\end{center}

Typical \mSR spectra in the normal and superconducting states at a low and a high hydrostatic pressure are shown in Fig. ~\ref{fig:msr spectra}. The data were fitted with the equations described in the Supplemental Material to extract the moments of the field distribution probed by the muons thermalized in the sample \cite{SM}. The temperature dependence of the spin depolarization rate $\sigma$ (proportional to the second moment of the field distribution) of muons stopping in the \cis sample and of the average internal field at different hydrostatic pressures are shown in Fig.~S1 and Fig.~S2 of the Supplemental Material, respectively.

 %


Typically $\sigma$ increases from $\sigma_{\mathrm n}$ in the normal state, to $\sigma=\sqrt{\sigma^2_{\mathrm{s}} + \sigma^2_{\rm n}}$ below $T_c$,
reflecting the inhomogeneous field distribution due to the decrease of the effective magnetic penetration depth $\lambda$ or, equivalently, to the increase of the effective superfluid density $\rho_s \equiv n_s/m^*$ when entering the vortex state ( $1 / \lambda(T)^2 \propto n_s(T)/m^*$, $n_s(T)$ density of supercarriers, $m^*$ effective mass). Fig.~S2 clearly shoes that $\sigma_{\mathrm{s}}(T)$ increases with pressure.

In an isotropic type-II superconductor with an hexagonal Abrikosov vortex lattice described by Ginzburg-Landau theory, the magnetic penetration depth $\lambda$ is related to $\sigma_{\mathrm{s}}$ by the equation ~\cite{Brandt03}:
\begin{equation}
\sigma_{\mathrm{s}}(T)[\mu{\rm s}^{-1}]=4.854\times10^4 (1 - b) \left[ 1 + 1.21(1 -\sqrt{b})^3 \right]\lambda(T)^{-2}[{\rm nm}^{-2}],
 \label{eq:Brandt_equation}
\end{equation}

Here $b=\left\langle B\right\rangle/B_{\rm c2}$ is a reduced magnetic field, with $\left\langle B(T)\right\rangle$ the first moment of the field distribution determined from the fit (see Supplemental Material and Fig.~S1) and $B_{\rm c2}(T)$ the second critical field determined by magnetoresistance and corrected for the small pressure effects according to Ref. \cite{Goh}. These terms in Eq. \ref{eq:Brandt_equation} take into account that the intervortex distance decreases with increasing applied field thus narrowing the field distribution. They are small in our case since in the relevant temperature range $\left\langle B\right\rangle \cong B_{applied} << B_{\rm c2}$.

\begin{figure}[htb]
\includegraphics[width=1.15\linewidth]{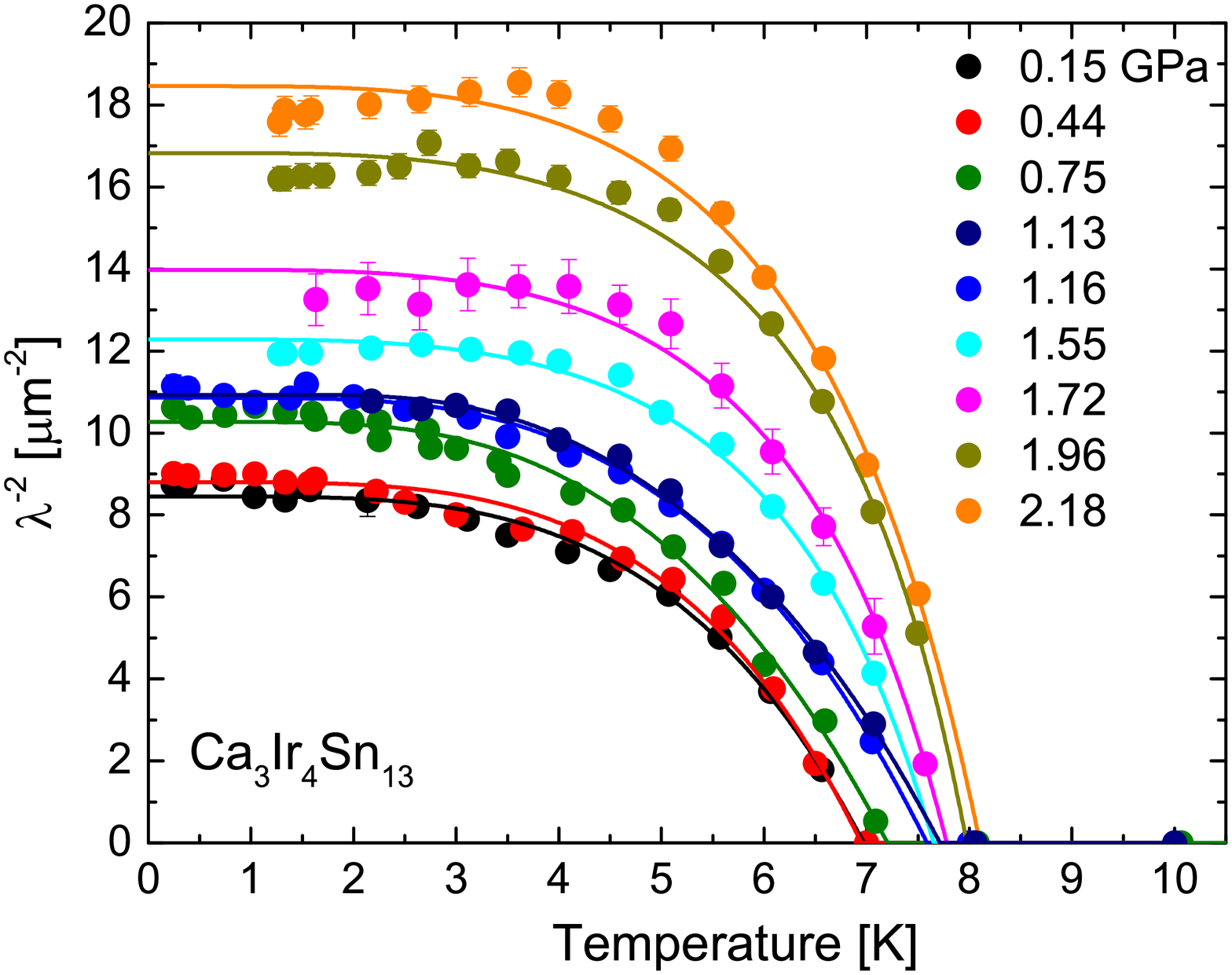}
\vspace{0cm}
\caption{Temperature dependence of the superfluid density in the vortex state of Ca$_{3}$\-Ir$_{4}$\-Sn$_{13}$, measured at various pressures after field cooling in 50 mT. The fit lines were obtained using the procedure explained in the text.}
\label{InvLambda2-vs-T}
\end{figure}

\begin{figure}[htb]
\includegraphics[width=0.9\linewidth]{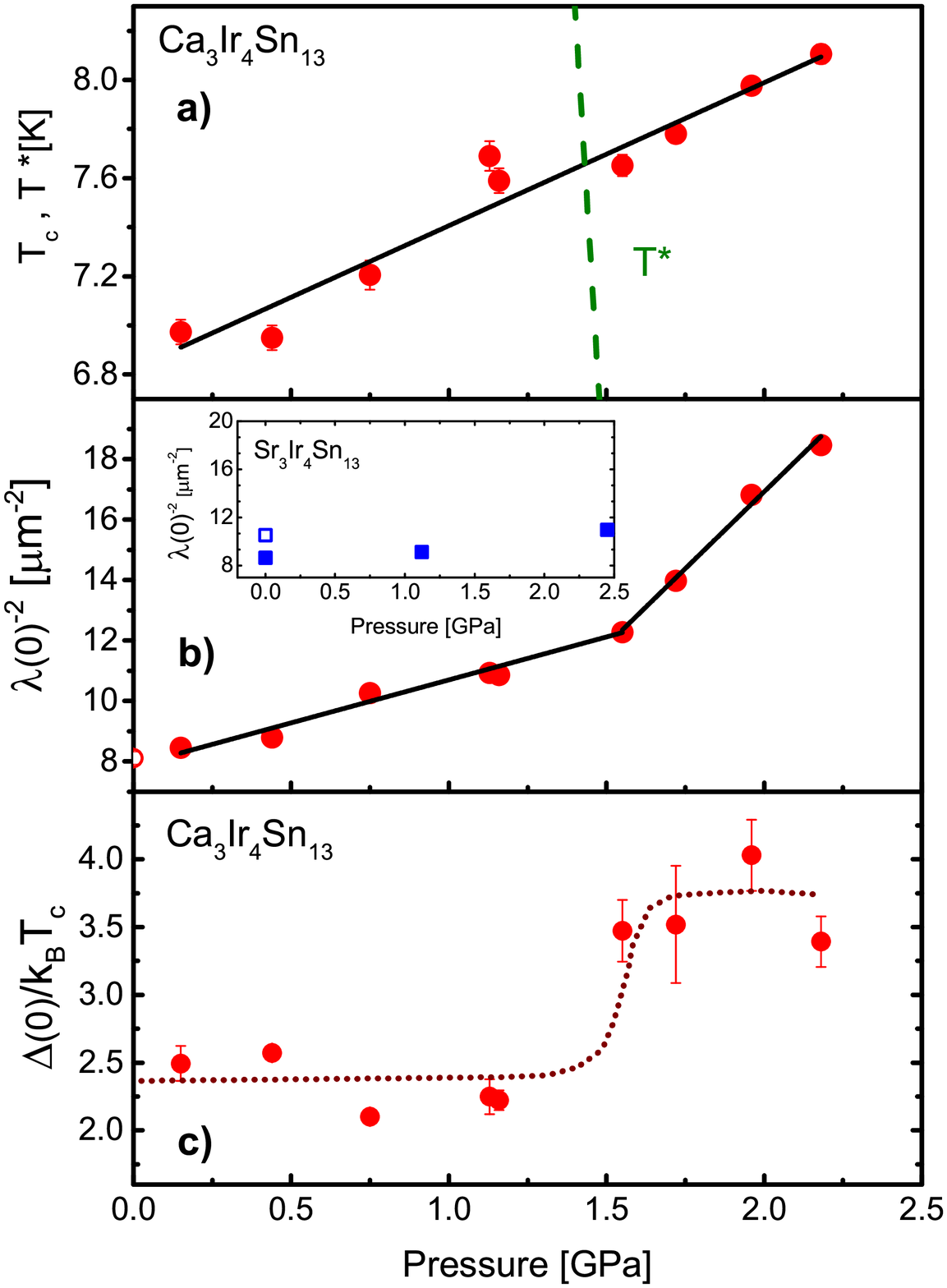}
\vspace{0cm}
\caption{a) Critical temperature versus applied pressure in \cis with linear fit. The dashed line indicates the $T^{*}$ pressure dependence, from \cite{Klintberg}. No anomaly is observable around the critical pressure $p_c \approx$ 1.6 GPa. b) Pressure dependence of the superfluid density of \cis at $T=0$,  proportional to $\lambda(0)^{-2}$ obtained from the fit shown in Fig. \ref{InvLambda2-vs-T}. Lines are linear fits in the two pressure regimes below and above the critical pressure. The inset shows the corresponding pressure range for \sis\!, which is offset with respect to \cis by $-5.2$ GPa. Open symbols are $p=0$ data from \cite{Biswas, Biswas14b}. c) Gap-to-\Tc ratio $R \equiv  \Delta(0) / (k_{\mathrm{B}} T_c)$ vs.\ pressure in \cis\!. The dashed line is to guide the eyes.}
\label{Par-vs-p}
\end{figure}

Figure \ref{InvLambda2-vs-T} shows the temperature dependence of $1 / \lambda^2$ taken at various pressures. The measurements at 1.13 and 1.16 GPa were performed with the MP35N and the CuBe cell, respectively. The results agree very well demonstrating the absence of systematic errors related to the cell type. Inspection of the figure clearly shows that the superfluid density increases with pressure and that this increase is more pronounced at pressures $p \gtrsim$ 1.7 GPa. For all pressures, there is no pronounced temperatures dependence at low temperatures, which is typical for a fully gapped $s$-wave superconductor, where $\Delta \lambda(T)= \lambda(T) - \lambda(0)$ decays exponentially; we find similar behavior for Sr$_{3}$\-Ir$_{4}$\-Sn$_{13}$. For all investigated pressures the data can be well fitted using in Eq. \ref{eq:Brandt_equation} the expression for a dirty superconductor with a single $s$-wave gap \cite{Tinkham04}:
\begin{eqnarray}
\frac{\rho_s(T)}{\rho_s(0)}=\frac{\lambda^2(0)}{\lambda^2(T)}= \frac{\Delta(T)}{\Delta(0)}\,\tanh\frac{\Delta(T)}{2 k_{\mathrm{B}} T},
\label{rho-vs-T}
\end{eqnarray}
where $\Delta(T)$ is the BCS superconducting gap. The dirty character of \cis is confirmed by our determination of various normal state and superconducting parameters, which are presented in the discussion section.
A useful parametrization of the BCS gap \cite{Muhlschlegel} is given in Ref.~[\onlinecite{Carrington}] by $\Delta(T)=\Delta(0)\tanh\{1.82[1.018(T_c/T-1)]^{0.51}\}$. Such a parametrization has been found to well represent the temperature dependence at any coupling strength \cite{Padamsee}. The fits yield the pressure dependence of the critical temperature $T_c$, of the superfluid density $1 /\lambda(0)^2$, and of the superconducting gap $\Delta(0)$.
The results are shown in Figs. \ref{Par-vs-p}a), \ref{Par-vs-p}b) and \ref{Par-vs-p}c), respectively.
The critical temperature increases almost linearly with increasing pressure. A linear fit yields $dT_c/dp =  0.592 (76)$ K/GPa with an intercept at $p=0$ of  6.79(6) K.
The smooth $T_{c}(p)$ dependence is in sharp contrast to the behavior of the superfluid density and of the gap to \Tc ratio as a function of pressure, where the changes cannot be simply related to pressure changes in \Tc.
The dependence of the superfluid density on pressure displays a pronounced change of slope above a pressure $p_c \approx$ 1.6 GPa.
 Above this value the slope is more than a factor of 3 higher than below: $d\lambda(0)^{-2}/dp = 2.84(18) \mu$m$^{-2}$/GPa for $p\leq p_c$ and $d\lambda(0)^{-2}/dp = 10.17(62) \mu$m$^{-2}$/GPa for $p\geq p_c$.
Equally remarkable is the jump of the $\Delta(0) / (k_{\mathrm{B}} T_c)$ ratio at $p_c$ from values around 2.2, typically of ternary stannide superconductors at $p=0$ \cite{Kase11} to values of about 3.7.
All these changes indicate the presence of a quantum critical point at $p_c \approx 1.6$ GPa, which, taking into account the uncertainties of a linear extrapolation, is close to the value, which was postulated from a linear extrapolation of the normal state values of $T^* (p)$ to $T=0$ \cite{Klintberg}.
Whereas part of the linear increase of $\lambda(0)^{-2}$ for $p < p_c$ can be related to the corresponding linear decrease of the mean free path of \cis \cite{Klintberg}, the pronounced changes at and above $p_c$ point to a profound modification of the electronic structure and of the superconducting interaction strength at $p_c$ and are strong indication of the presence of a quantum critical point in the superconducting phase.
Similar pressure measurements of \sis for $0.15 <p< 2.5$ GPa corresponding to values between $-5.1$ GPa and $-2.7$ GPa of the \csis phase diagram, far away from $p_c$, do not exhibit pronounced or anomalous pressure effects (inset of Fig. \ref{Par-vs-p}b).
The total amplitude of the Gaussian signals of the \mSR spectra, which is a measure of the volume fraction of the superconducting phase does not show any pressure dependence and has the maximum possible value indicating a full volume superconducting phase below and above the quantum critical point.

\section{Discussion}\label{sec:Discussion}

\subsection{Pressure dependence of superconducting parameters}

Application of pressure allows to uncover superconductivity and plays an important role to obtain information about its microscopic mechanism.
For instance the dependence of \Tc on pressure for MgB$_2$ indicated a mediating electron-phonon coupling \cite{Tomita01}.
Recently, in Bi$_2$Se$_3$ an anomalously saturating $p$ dependence of \Tc (for $p > 30.0$ GPa) suggested the presence of an unconventional pressure-induced superconducting pairing state in this topological insulator \cite{Kirshenbaum13} and pressure-induced reversal of the dependence of \Tc on pressure in
KFe$_2$As$_2$ indicated a pairing state transition \cite{Tafti}. Investigations have mainly focussed on the study of $T_{\rm c}(p)$ \cite{Schilling01}.
For the most simple metals and 
in the majority of the prototypical BCS superconductors, superconductivity is reduced
under pressure $dT_{\rm c}/dp < 0 $.
However, even in elementary superconductors the effects of pressure can be complicated \cite{Buzea05}.
The variation of the critical temperature with pressure can be generically obtained by differentiating the McMillan expression for strongly coupled superconductors with respect to \Tc  \cite{MacMillan68,Tomita01,Khasanov04}.

\begin{equation}
\label{eq: pressure shift}
\frac{d \ln T_{c}}{d p} = -(2A-1) \frac{d \ln \langle \omega_{ln} \rangle}{d p} + A \frac{d \ln \eta}{d p}
\end{equation}
where $A=1.04 \lambda_{ep} [1+0.38\mu^{\ast}]/[\lambda_{ep}-\mu^{\ast}(1+0.62\lambda_{ep})]^{2}$ is a function of the electron-phonon coupling constant $\lambda_{ep}$ and of the Coulomb pseudopotential $\mu^{\ast}$, $\langle \omega_{ln} \rangle$ is the logarithmic averaged phonon frequency, $\eta \equiv N(E_{F})\langle I^{2} \rangle$ is the Hopfield parameter with $\langle I^{2} \rangle$ the electronic matrix element of the electron phonon interaction averaged over the Fermi surface  and $N(E_{F})$ the density of states at the Fermi level.

 To the pressure dependence of \Tc there are phononic and electronic contribution. In conventional phonon mediated superconductors the phonon stiffening with pressure leads to a decreasing critical temperature, albeit, close to a structural transition, phonon softening of particular modes can occur. For a BCS superconductor the superfluid density and hence $\lambda(0)$ is pressure independent as well as the ratio $\Delta(0) / (k_{\mathrm{B}} T_c)$. This is the case for instance of RbOs$_2$O$_6$
 \cite{Khasanov04}. Also in the multigap superconductor MgB$_2$, which is a moderately strong electron-phonon mediated superconductor, the two gap-to-\Tc ratios are practically pressure independent and only a small pressure effect of $\lambda(0)$ has been observed \cite{DiCastro05}.
No or small pressure dependence of $\lambda(0)^{-2}$ has been reported also in some unconventional superconductors, such as the electron doped infinite layer cuprate Sr$_{0.9}$La$_{0.1}$CuO$_2$ \cite{DiCastro09} or the nearly optimally hole doped YBaCu$_3$O$_{7-\delta}$ \cite{Maisuradze11}.

In the \csis system, the positive variation of the critical temperature with pressure is a clear indication of the importance of the electronic contribution, such as a change of $N(E_{F})$.
The observed pressure induced enhancement of the superfluid density is very high, with a relative variation between the lowest (0.15 GPa) and the highest (2.18 GPa) measured pressures $\Delta \lambda(0)^{-2}/ \lambda(0)^{-2} \simeq 240 \%$. This is unusual for a BCS superconductor.
\subsection{Phase Diagrams and Uemura plot}

The pronounced increase setting in above 1.6 GPa can be understood if put in relation with the structural quantum phase transition and the related CDW suppression at $p_c$.
Superconducting and CDW gap are often antagonistic to each other. The CDW gapping inhibits superconductivity by decreasing the area of the Fermi Surface where the superconducting gap can open and by reducing the number of available carriers. This is apparent in various physical properties of the \csis system. Optical spectroscopy measurements across the structural/CDW phase transition on single-crystal samples of Sr$_3$Ir$_4$Sn$_{13}$ have determined from the Drude components a reduction of the plasma frequency  from $\omega_p \approx 30530$ cm$^{-1}$ above to $\omega_p \approx 25750$ cm$^{-1}$ below $T^*$ \cite{Fang14}.
If the effective mass of the itinerant carriers remains unchanged, this means that roughly 29\% of the carriers are lost below $T^*$. In \cis the temperature dependence of the Hall coefficient shows a significant reduction of carriers below $T^* $ \cite{Wang}. A reduction of the electronic density of states at E$_F$ in the CDW phase has also been predicted by DFT calculations of the electronic structure of \sis \cite{Klintberg, Tompsett14}.


Superconductivity in metallic ternary stannide compounds has been analysed within the theory of strong coupling superconductivity, where the gap
can be expressed as a function of the logarithmic averaged phonon frequency
${{\omega }_{\ln }}\equiv \exp \left( \frac{2}{{{\lambda }_{ep}}}\int\limits_{0}^{\omega }{\ln \omega {{\alpha }^{2}}F(\omega )dln\omega } \right)$ and $x \equiv \frac{\omega_{\ln}}{T_{\rm c}}$

\begin{eqnarray}
\Delta(0)/k_{\rm B}T_{\rm c}= 1.768 \left[ 1+ \frac{12}{x^2}\ln\frac{x}{2} \right]
\label{Delta}
\end{eqnarray}

For \sis and \cis specific heat and magnetic susceptibility measurements yield $\omega_{ln}$= 73 K and 56 K, respectively.
The reduced values of $\omega_{ln}$ as compared to that of the Debye temperature ($\omega_D$=184 K and 218 K, respectively)
indicate the importance of the low-energy phonon modes for the superconductivity in the ternary stannide compounds \cite{Kase11, Hayamizu}.
The pronounced increase of the gap-to-\Tc ratio at $p_c$ however, cannot be simply understood within Eq. \ref{Delta}, which only predicts a maximum value for the ratio of 2.77, well below our observation (see Fig. \ref{Par-vs-p}c)).
Since our data exclude the presence of spin fluctuations or magnetic glue at any pressure, it is natural to attribute the observed effect to the particular role of phonons associated with the structural QPT and their synergetic effects with concomitant changes of the electronic spectrum related to the CDW suppression.
The DFT calculations by Tompsett have identified low energy phonon modes at the ${\bf X}$ and ${\bf M}$ points as responsible for the structural transition. A particular role appears to be played by the phonon mode at the ${\bf M}$ point which corresponds to a breathing of the Sn$_{12}$ cages \cite{Tompsett14}.
These phonon modes go soft at $T^*$. Softening of a low energy phonon mode associated with the Sn$_{12}$ breathing mode has been also found by inelastic neutron scattering \cite{Mazzone}. When the transition temperature is driven to 0 K by pressure, the softening occurs at this temperature, giving rise to additional low-lying phonon modes, which can be excited at low temperatures.
The effect of these additional phonon modes on the superconducting strength can be estimated using a model originally proposed for the fullerides, which take into account electron coupling to phonons with very large differences in frequency \cite{Mazin93}. Such a model is able to produce a gap-to-\Tc ratio of 5 well encompassing the observed increase to $\sim$ 3.7 at $p_c$ and is able to explain the transition from
intermediate strong to very strong superconductivity at pressures $p>p_c$ \cite{Mazin93}.
This result suggests that the low energy phonons are not only responsible for the structural transition but also
drive and boost the competing superconducting phase, which reaches its maximum \Tc at $\sim$ 4.0 GPa.
A similar mechanism could be at the origin of the superconducting dome in the CDW superconductor 1$T$-TSe$_2$ when the CDW transition is suppressed by pressure within a behavior
also reminiscent of a quantum critical point \cite{Kusmartseva}.
However, we cannot exclude that a role may be also played by charge density fluctuations on approaching the QCP.

\begin{figure}[tph]
\includegraphics[width=1.15\linewidth]{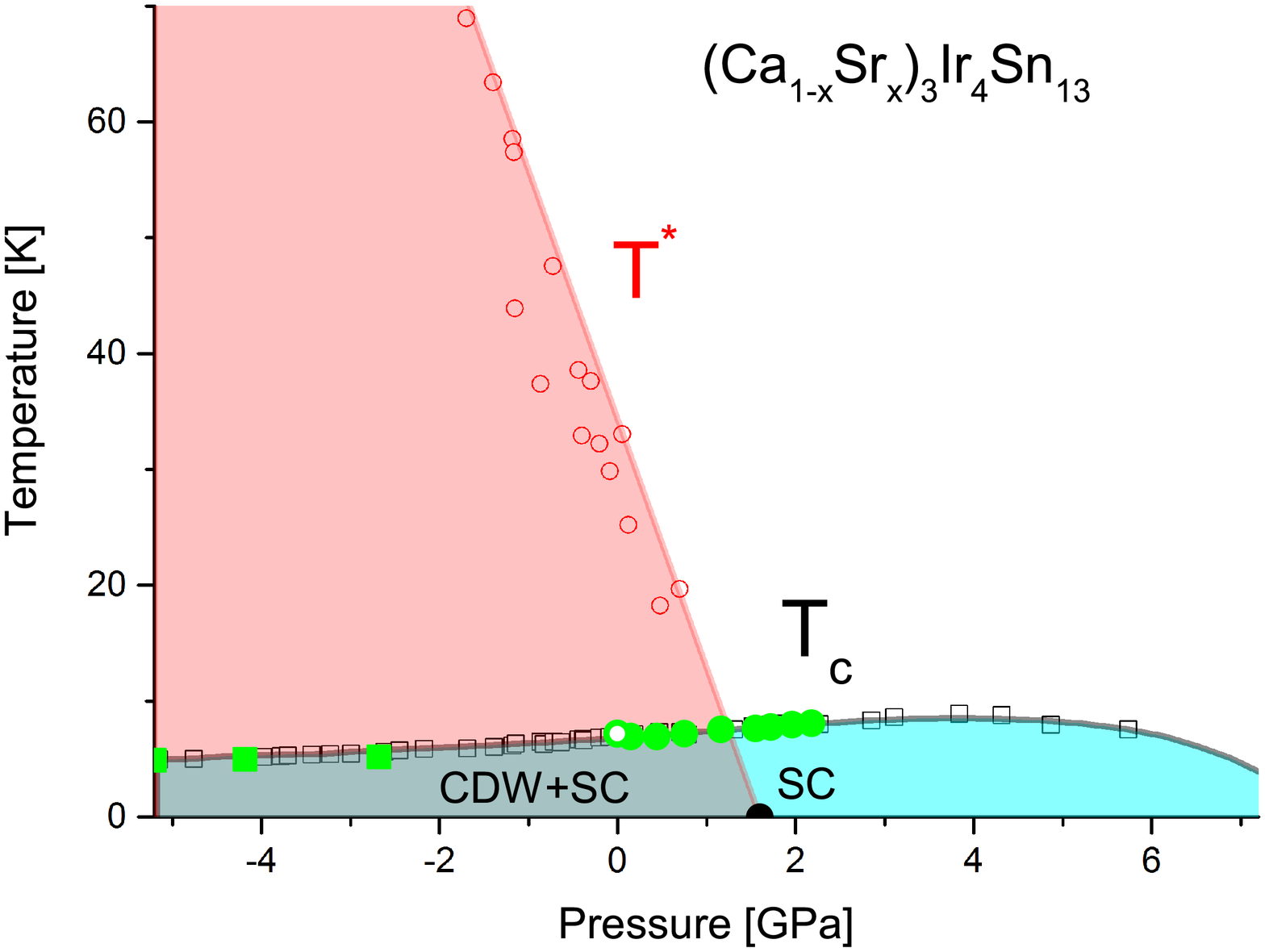}
\vspace{-0cm}
\caption{Low temperature phase diagram of \csis proposed by the present \mSR investigations of the microscopic superconducting parameters. They indicate that there are two superconducting phases, one coexisting with CDW and the other pure. They are separated by a critical point at $\approx $ 1.6 GPa, indicated by a black dot. Green circles and squares: superconducting transition temperatures of \cis and \sis under pressure obtained from the \mSR data. Open green circle is a point at $p=0$ from \cite{Biswas14b}. Open squares are \Tc values from AC susceptibility and resistivity measurements and open circles $T^*$ values from resistivity measurements, data from Ref. \cite{Klintberg}.}
\label{PhaseDiagram}
\end{figure}

\begin{figure}[tph]
\includegraphics[width=1.0\linewidth]{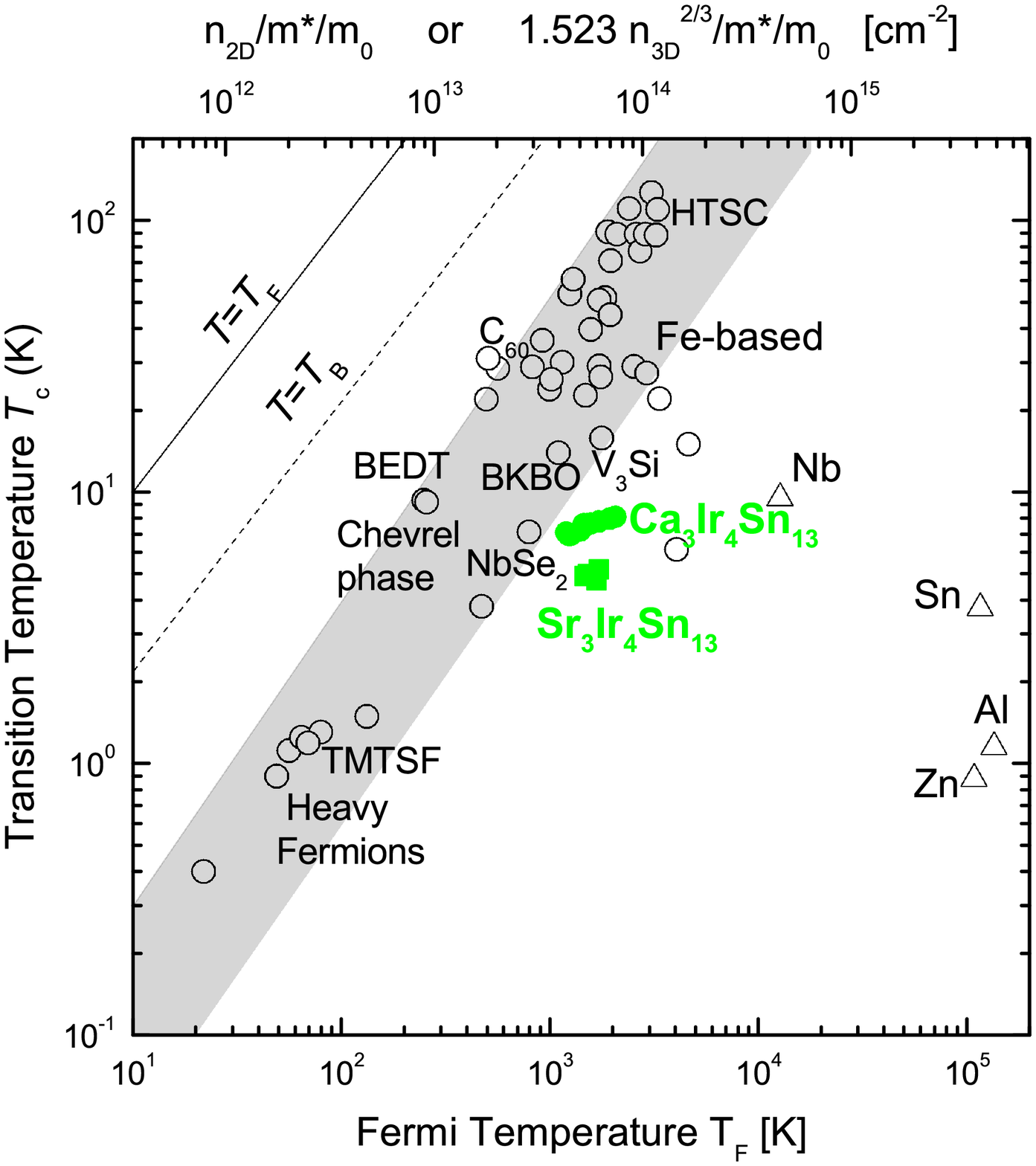}
\vspace{-0cm}
\caption{Uemura plot. Green points: The superconducting transition temperature \Tc vs the effective Fermi temperature $T_F$ for \cis and \sis at various pressures. The points are
plotted in the \Tc$-T_F$ diagram together with data of different families of superconductors (plot adapted from \cite{Khasanov08,Hashimoto}).
Top scale: $T_F$ expressed in terms of paired carriers density and effective mass, evaluated from the $\lambda(0)^{-2}$ measurements:
1.52 $n_s^{2/3}/\frac{m^{\ast}}{m_e}$ for a 3D system such as \csis and $n_s/\frac{m^{\ast}}{m_e}$ for 2D systems.
The unconventional superconductors are generally considered to fall within a band, indicated by the gray region in the figure.
Conventional elemental superconductors lie on the right side of the diagram. The dashed line corresponds to the Bose-Einstein condensation temperature $T_B$.}
\label{UemuraPlot}
\end{figure}

The pronounced increase of $\rho_s$ and $\Delta(0) / (k_{\mathrm{B}} T_c)$ when $T^*$ is driven to zero clearly indicates that below $p_c$ the CDW gap is competing with the superconducting gap and that only its complete suppression allows superconductivity to fully develop. However,  from the amplitude of the $\mu$SR signal we can deduce that, while competing for the same electrons, the two order parameters coexist at nanoscale with the superconducting volume faction remaining 100\% over the entire pressure range. Overall our results indicate that the \csis system has a quantum critical point at $\approx$ 1.6 GPa, separating two different superconducting states, a low pressure one coexisting with a CDW and a pure superconducting one at higher pressures, resulting in a phase diagram as shown in Fig.~\ref{PhaseDiagram}.

From the data we can also estimate the increase of available carriers associated with the closing of the CDW gap.  As a measure we take the relative difference between $\lambda(0)^{-2}$ of \cis at 1.16 GPa where superconductivity is coexisting with the CDW phase at all temperatures and $\lambda(0)^{-2}$ at 2.18 GPa where the CDW is completely suppressed. The relative increase in superfluid density is 68(4)\%.  DFT calculations of the electronic structure  of \sis with the generalized gradient approximation (GGA) have predicted a $\sim$ 34\% increase of $N(E_F)$ going rising the temperature above $T^*$ \cite{Klintberg}. Within the simplified assumption of a free electron gas model, where
$N(E_F)\propto n^{-1/3}$ this corresponds to a $\sim$ 140\% increase of charge carriers. This number appears reasonable if one takes into account that changes in $\lambda(0)^{-2}$ also reflect changes of the effective mass $m^{\ast}=m_b (1 + \lambda_{int}$), which we may expect to increase by entering the very strong coupling regime above $p_c$.


The difference by almost a factor of two of the gap amplitudes in the two superconducting phases is in agreement with the dependence of the BCS ratio on the CDW gap and the degree of Fermi surface gapping as calculated in \cite{Garovich02}, where it was shown that the electron-hole gapping in the CDW-superconducting phase may lead to a reduction of the BCS ratio by as much as one half.
Our data also show that while the gap amplitude $\Delta(0)$ strongly depends on the suppression of the CDW, the temperature dependence of the original BCS expression is not affected by the presence of the CDW and is independent from the coupling strength.
On the other hand, comparison of Fig. \ref{Par-vs-p}a) with Fig. \ref{Par-vs-p}b) and c) shows that \Tc is much less sensitive to the quantum critical point than the superfluid density and the coupling strength.
This reflects the fact that, whereas closing the CDW gap makes electronic states at the Fermi surface available for superconductivity and directly increases $n_s$, the electronic contribution to $dlnT_c/dp$ as expressed by Eq.  \ref{eq: pressure shift} only depends logarithmically on $N(E_F)$.

Whereas, BCS superconductors did not show any pressure effects on the superfluid density, more or less pronounced effects have been observed in unconventional superconductors.
In layered cuprates, in the presence of charge reservoir layers, pressure induced charge transfer to the CuO$_2$ planes is an important mechanism besides changes of the effective pairing interaction to increase $\rho_s$.
A large transfer of holes from the double chains to the CuO$_2$ planes has been observed in YBa$_2$Cu$_4$O$_8$, which also displays large pressure derivative of \Tc
and large increase of superfluid density $\frac{\Delta (\lambda(0)^{-2})}{\lambda(0)^{-2}} = 42 \%$ at 1.0 GPa pressure. Here, however, about two thirds of the effect have been attributed to a reduction of the effective mass \cite{Khasanov05}.
 Pronounced pressure dependence of $\rho_s$ or pressure induced superconductivity has been often found in unconventional systems with magnetic phases and where magnetic fluctuations are the most probable candidates for an attractive interaction. Examples are   
 (pressure induced) superconductivity in the antiferromagnetic phase of the non-centrosymmetric heavy fermion CeRhSi$_{3}$ \cite{Kimura}
, in the ferromagnetic phase of UGe$_2$ \cite{Saxena}. In the prototypical heavy Fermion superconductor CeCoIn$_5$, the increase of $\rho_s$
under pressure corresponds to about a  doubling of the supercarrier number density $n_s$ between 0 and 1.0 GPa. The (smooth) increase was related to the possible presence of a quantum critical point \cite{Howald}.

The phase diagram of \csis reminiscent of unconventional superconductors,  the pronounced pressure dependence of $\rho_s$, and the jump of gap and coupling strength at the quantum critical point raise the question about a possible unconventional character of this compound.
Unconventional superconductivity is often found in strongly correlated electron systems, displays order parameter with gaps which are anisotropic and displaying sign reversal between different Fermi surface pockets
or is related to intrinsic magnetism, connected to competing magnetic phases  or display broken symmetries, such as invertion or time reversal symmetry.
None of these characteristics is found in this system, which has an $s$-wave order parameter all over the phase diagram and attractive interaction where phonons appear to play a dominant role.

\cis appears to have moderately heavy carriers and weak correlation \cite{Wang15,Wang}.
The dimensionless ratio of the thermopower (Seebeck coefficient) $S/T$ to the specific heat term $q= \frac{N_{Av}e S}{T \gamma}$ provided a carrier density $\left| q^{-1} \right| \cong 11.8(8)$ per f.u., corresponding to high DOS per volume \cite{Wang}.
Note that from the Sommerfeld coefficient for \cis $\gamma_{meas}=39$ mJ/K$^2$ mol \cite{Wang} we get a similar carrier density n$_e$ = 5.3 $ \times  10^{22}$ cm$^{-3}$ which gives
a BCS coherence length $\xi_0=  \hbar^2 (3 \pi^2 n_e)^{\frac{1}{3}} / [m_e(1+\lambda_{e-ph}) \pi \Delta(0)] \sim 80$ nm (where we used for the electron-phonon coupling $\lambda_{e-ph}=1.34$ (from \cite{Hayamizu}) and $\Delta(0)$=1.51 meV).
Using the value of the residual resistivity above $T_c$ ($\rho_n = 79 \mu\Omega\,$cm) we estimate the mean free path $\ell= \hbar (3 \pi^2)^{\frac{1}{3}}/ (n_e^{\frac{2}{3}} \rho_n e^2)$ = 1.14 nm $ << \xi_0$.
The dirty character of the superconductor is consistent with our analysis of $\lambda(T)$.

The effective coherence length obtained from $B_{c2}(0)=5.4 $T, determined from the well-known WHH formula  $B_{c2}(0)=-0.693\, T_c\, \frac{dB_{c2}}{dT}$,
is $\xi_{\mathrm{GL}} = \sqrt{\frac {\Phi_0}{2\pi B_{c2}}} \cong $ 7.8 nm, where $\Phi_0 = 2.07 \times 10^{-15}$ T\,m$^2$ is the quantum of magnetic flux.
The good agreement of $\xi_{\mathrm{GL}}$ with the value obtained from $\sqrt{\xi_0 \ell} \simeq 9.5 $ nm, gives us further confidence about the correct estimate of these electronic parameters.
The low value of $\ell / \xi_0 = 0.014$ means that in \cis the density of paired electrons is reduced to $n_s \thickapprox n_e \frac{\ell}{\xi_0} = 7.4 \times 10^{20} $ cm$^{-3}$.
From our determination of the effective magnetic penetration depth $\lambda$, we can alternatively estimate the density of paired electrons to be $n_s = \frac{m(1+\lambda_{e-ph})}{\mu_0 e^2 \lambda^2} \simeq 5.36 \times 10^{20}$ cm$^{-3}$. This number is in good agreement with the above estimate, which is based on a carrier density expressed in terms of a simple free electron model, and reconfirms the applicability of the BCS theory.



To place \csis in the context of other superconductors, we plot in Fig.~\ref{UemuraPlot} the measured points in the Uemura graph of \Tc versus effective Fermi temperature, which is often used to define the
character of unconventionality of a superconductor \cite{Hashimoto, Uemura04}.
The green symbols correspond to different pressures for \cis studied here, the others points are taken from \cite{Khasanov08,Hashimoto}.
Remarkably, plotting the values of \Tc as a function of the effective Fermi temperature which for a 3D systems $k_B T_F=\frac{\hbar^2}{2}(3\pi^2)^{2/3}\frac{n_s^{2/3}}{m^{\ast}}$, where n$_s$ is the value determined above from $\lambda$ we observe that \cis lies at the border of that part of the diagram where cuprates and iron based superconductors and other unconventional superconductors are found.
A similar result is obtained if we express the superfluid density in term of the muon spin relaxation rate $\sigma(0) \propto \lambda(0)^{-2} \propto \rho_s(0)$ as in the original Uemura plot.
This behavior shows that a conventional electron phonon mediated cubic superconductor is not a criterium of exclusion from this part of the plot. The vicinity to unconventional superconductors
is possibly related to the competing CDW state or to the presence of a quantum phase transition. Remarkably the investigated system is placed in the Uemura plot close to V$_3$Si, another phonon-mediated superconductor with a nearby structural instability.

Recently, it has been reported that the series (Ca$_{1-x}$,Sr$_x$)$_{3}$\-Rh$_{4}$\-Sn$_{13}$ has a very similar phase diagram with a quantum critical point which should appear at ambient pressure
if the Ca fraction is 0.9 \cite{Goh15}. This opens the possibility to study the interplay of superconductivity, CDW and quantum criticality over a broader interval of pressures up and beyond the region where superconductivity is strongest and the critical temperature reaches its maximum.

{\bf Note added}. After submission of this paper strong coupling superconductivity has been also reported near a structural quantum critical point of (Ca$_{1-x}$,Sr$_x$)$_{3}$\-Rh$_{4}$\-Sn$_{13}$
 \cite{Yu15}.

\section{Acknowledgement}

Work at PSI was supported by the Swiss National Science Foundation.
Work at Brookhaven is supported by the U.S. DOE under Contract No. DE-SC00112704.

%
%

\end{document}



\title {Strong enhancement of s-wave superconductivity near a quantum critical point of Ca$_3$Ir$_4$Sn$_{13}$}

\author{P.~K.~Biswas}
\affiliation{Laboratory for Muon Spin Spectroscopy, Paul Scherrer Institute, CH-5232 Villigen PSI, Switzerland}
\author{Z. Guguchia}
\affiliation{Laboratory for Muon Spin Spectroscopy, Paul Scherrer Institute, CH-5232 Villigen PSI, Switzerland}
\author{R.~Khasanov}
\affiliation{Laboratory for Muon Spin Spectroscopy, Paul Scherrer Institute, CH-5232 Villigen PSI, Switzerland}
\author{M.~Chinotti}
\affiliation{Laboratory for Muon Spin Spectroscopy, Paul Scherrer Institute, CH-5232 Villigen PSI, Switzerland}
\author{L. Li}
\affiliation{Condensed Matter Physics and Materials Science Department, Brookhaven National Laboratory, Upton, New York 11973, USA}
\author{Kefeng~Wang}
\affiliation{Condensed Matter Physics and Materials Science Department, Brookhaven National Laboratory, Upton, New York 11973, USA}
\author{C.~Petrovic}
\affiliation{Condensed Matter Physics and Materials Science Department, Brookhaven National Laboratory, Upton, New York 11973, USA}
\author{E.~Morenzoni}
\email[Corresponding author: ]{elvezio.morenzoni@psi.ch}
\affiliation{Laboratory for Muon Spin Spectroscopy, Paul Scherrer Institute, CH-5232 Villigen PSI, Switzerland}
%

%


\maketitle

\newpage
\section{Supplementary Material}\label{sec:Supplement}
\setcounter{figure}{0}
\renewcommand{\thefigure}{S\arabic{figure}}

\noindent
\textbf{Methods}

From the TF-$\mu$SR spectra we can determine the second moment of the magnetic field distribution in the vortex state and from this the magnetic penetration depth. The distribution $p(B)$ can be well represented by a multi-component Gaussian curve, which corresponds having the sample contribution to the muon time spectra fitted to a sum of $N$ Gaussian components ~\cite{Weber93,Maisuradze09}.
We first performed a full analysis with one Gaussian and then added an additional Gaussian, checking whether the $\chi^2$ of the fit improves significantly.
We found that with the CuBe cell slightly better fits are obtained at low temperatures ($T < 6$ K) with $N$=2.  At higher temperatures and for the data obtained with the MP35N cell a single Gauss fit well reproduces the data.
As mentioned previously, a fraction of the $\mu$SR asymmetry signal originates from muons stopping in the pressure cell wall, which is taken into account by the term $A_{\rm cell}(t)$. The \mSR asymmetry
spectra are fitted by following expression:
\begin{eqnarray}
A(t)=e^{-\sigma_{n}^2 t^2/2}\sum_{i=1}^{N}A_{i}e^{-\sigma_{i}^2 t^2/2}\cos{(\gamma_{\mu}B_{i}t+\varphi)}+A_{\rm cell}(t).
\label{Asy}
\end{eqnarray}

The second moment of the multi-Gaussian internal field distribution is then given by:

\begin{eqnarray}
\langle\Delta B^2\rangle=\frac{\sigma_{sc}^2}{\gamma_{\mu}^2} = \sum_{i=1}^{N} && \, \frac{A_i}{A_1+A_2+\ldots+A_N} \nonumber \\
&&\,\times \left[(\frac{\sigma_{i}}{\gamma_{\mu}})^{2}+(B_i-\langle B \rangle)^2 \right],
\label{Secondmoment}
\end{eqnarray}
with
\begin{eqnarray}
\langle B\rangle=\sum_{i=1}^{N}\frac{A_i B_i}{A_1+A_2+\ldots+A_N}.
\label{Firstmoment}
\end{eqnarray}
Here $\varphi$ is the initial muon spin phase, while $A_i$, $\sigma_{i}$, and $B_i$ are the amplitude, relaxation rate and first moment of the internal field of the $i$-th Gaussian component, respectively. $\sigma_{\rm n}$ is a (small) contribution to the field distribution arising from the nuclear moments. It is independent of temperature and pressure and was determined well above $T_c$.
The CuBe cell (used for the pressures 0.15, 0.44, 0.75 and 1.16 GPa) contribution is given by:
\begin{equation}
\label{eq: CuBe background}
A_{\rm cell}(t) = A_{\rm CuBe} e^{-\sigma_{\rm CuBe}^2 t^2/2} \cos{(\gamma_{\mu}B_{\rm CuBe}t+\varphi)}
\end{equation}
where $A_{\rm CuBe}$  the initial asymmetry, $\sigma_{CuBe}$ the relaxation of the pressure cell and $B_{\rm CuBe}$ the average field sensed by the muons stopping in the pressure cell.
The main contribution of the pressure cell relaxation is due to nuclear moments in CuBe, which amounts to a value of $0.26$ $\mu s^{-1}$.

The MP35N cell (used for the pressures 1.13, 1.55, 1.72, 1.96 and 2.18 GPa) has a dynamic electronic contribution in addition to the nuclear magnetic moments contribution.
It can be fitted by
\begin{equation}
\label{eq: MP35N background}
A_{\rm cell}(t) = A_{\rm MP35N} e^{-\sigma_{\rm MP35N}^2 t^2/2} e^{-\lambda_{\rm MP35N}(T)t} \cos{(\gamma_{\mu}B_{\rm MP35N}t+\varphi)}
\end{equation}
with $\sigma_{\rm MP35N}=0.25 \mu s^{-1}$ and the electronic contributions $\lambda_{\rm MP35N} (T)$ approximately constant and small ($\approx 0.05-0.1 \mu s^{-1}$ above $\sim 1$ K), but strongly temperature dependent below and rising to $\sim 0.6 \mu s^{-1}$ at the lowest temperature.
The average field in the cells is very close to the applied field. However, with the sample in the superconducting state, the associated diamagnetism leads to a magnetic field being induced around the sample, which could in principle produce an additional broadening and a decrease of the average field proportional to $B_{\rm app} - \langle B\rangle$ \cite{Maisuradze11}. We also fitted the data taking this effect into account, but did not find significant differences from the simpler fit assuming the Gaussian broadening and the average field in the cells temperature and pressure independent.The very good agreement of the temperature dependence of the superfluid density and fit parameters obtained at 1.13 and 1.16 GPa (Fig. 2 and 3), which were measured with the two different cells types, shows the consistency of the data analysis.

The temperature dependence of the MP35N cell contribution makes a detailed determination of the superfluid density below 1K difficult. However this does not change the result of dominant $s$-wave character of the order parameter. Any other symmetry of the order parameter with nodes is not consistent with the shape of $1/\lambda^2(T)$ for $T \gtrsim$ 1 K.  The appearance of very small secondary gaps at very low temperature cannot be excluded a priori; however, their contribution to the superfluid density would be negligible, otherwise they would affect the higher temperature region as well.

\newpage
\noindent
\textbf{Supplementary figures}

\begin{figure}[tbh]
\centering
\includegraphics[width=0.8\textwidth]{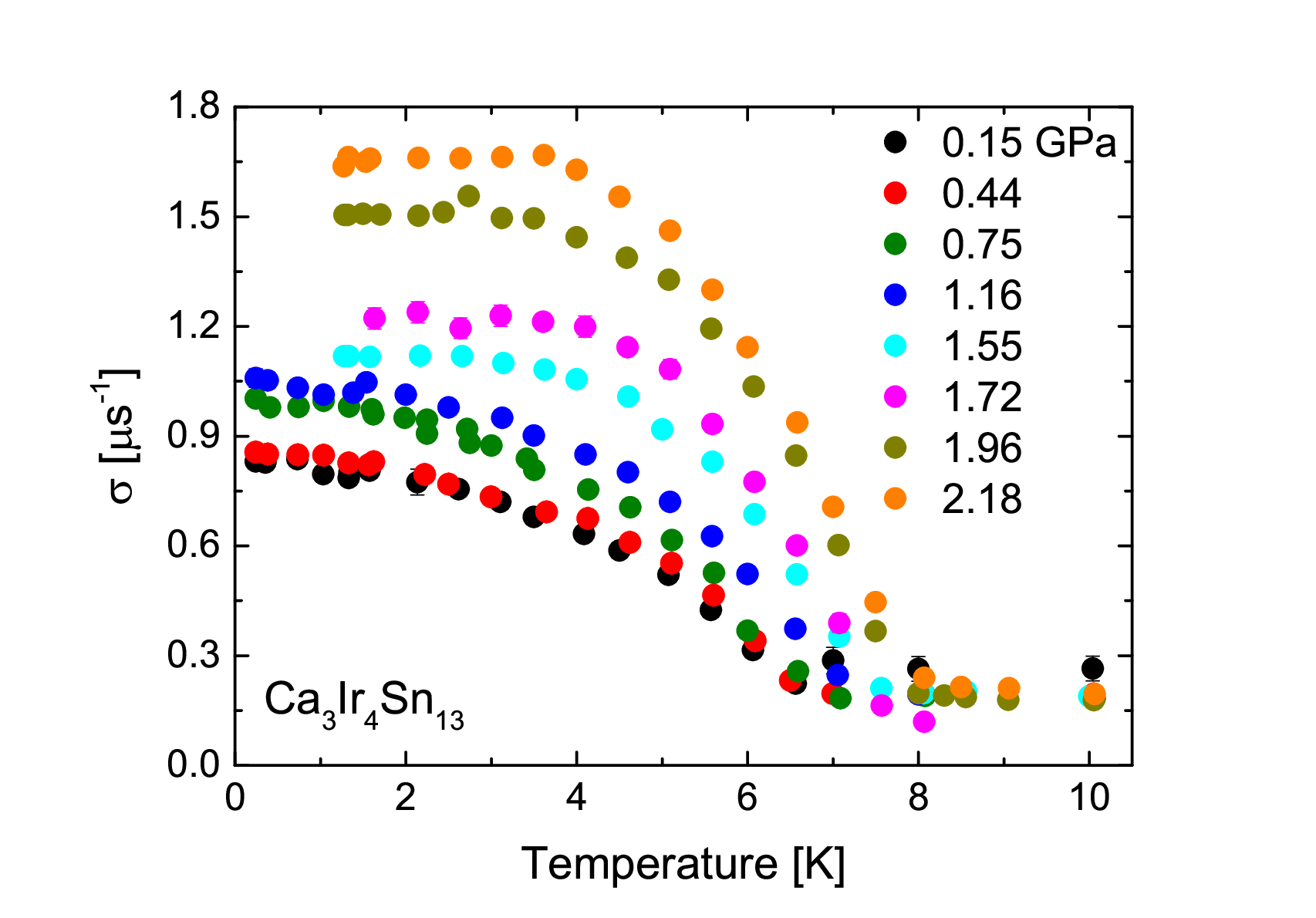}
\caption{\label{fig:sigma-vs-T-p} Temperature dependence of the muon spin depolarization rate $\sigma$ in the vortex state of \cis, measured at different pressures after cooling in 50 mT.
$\sigma$, which is directly proportional to the second moment of the field distribution has been obtained by fitting the $\mu$SR time spectra following the procedure described in the Methods section.}
\end{figure}

\begin{figure}[tph]
\includegraphics[width=0.8\linewidth]{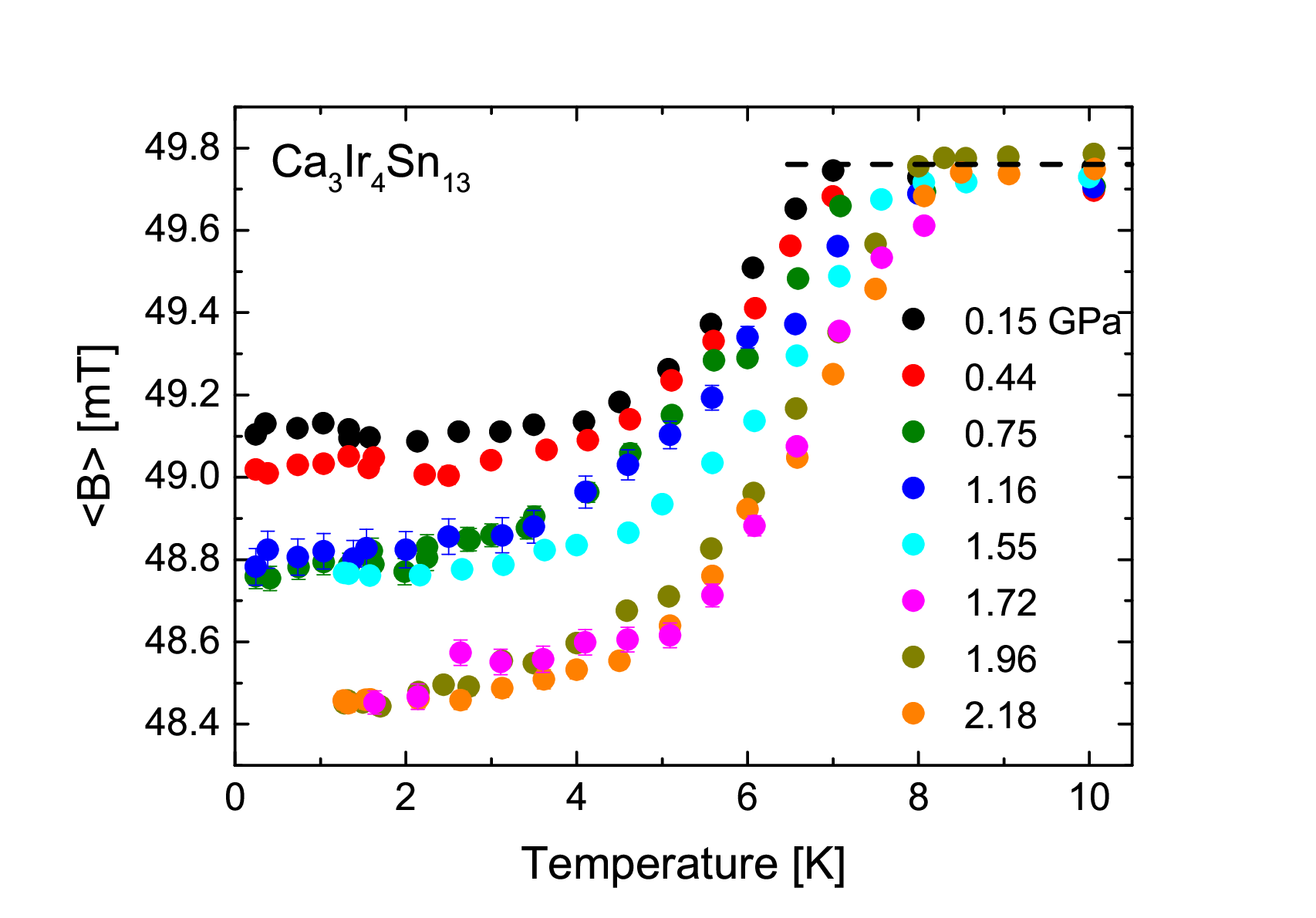}
\vspace{0cm}
\caption{ Temperature dependence of the average internal field (first moment of the field distribution) in the \cis sample for different pressures. The applied field is indicated by a dashed line. The diamagnetic shift in the vortex state is clearly visible. The exact value of the shift depends on demagnetizing corrections, which may be slightly pressure dependent.}
\label{IntField-vs-T-p}
\end{figure}

\pagebreak
%
%